\documentclass[aps,prb,final,twocolumn,superscriptaddress]{revtex4-1}

\usepackage{amsmath}
\usepackage{braket}
\usepackage{graphicx}
\usepackage{epsfig}
\usepackage{natbib}

\DeclareMathOperator{\Tr}{Tr}

\begin{document}


\title{Thermal fluctuations in the conical state of monoaxial helimagnets}


\author{Victor Laliena}
\email[]{laliena@unizar.es}
\affiliation{Instituto de Ciencia de Materiales de Arag\'on 
(CSIC -- Universidad de Zaragoza) \\ C/Pedro Cerbuna 12, 50009 Zaragoza, Spain}
\author{Yusuke Kato}
\affiliation{Department of Basic Science, The University of Tokyo, Meguro,
Tokyo 153-8902, Japan}
\author{Germ\'an Albalate}
\affiliation{Instituto de Ciencia de Materiales de Arag\'on 
(CSIC -- Universidad de Zaragoza) \\ C/Pedro Cerbuna 12, 50009 Zaragoza, Spain}
\author{Javier Campo}
\email[]{javier.campo@csic.es}
\affiliation{Instituto de Ciencia de Materiales de Arag\'on 
(CSIC -- Universidad de Zaragoza) \\ C/Pedro Cerbuna 12, 50009 Zaragoza, Spain}




\date{\today}

\begin{abstract}
The effect of thermal fluctuations on the phase structure of monoaxial helimagnets 
with external magnetic field parallel to the chiral axis is analyzed by means of a 
saddle point expansion of the free energy. The phase transition that separates
the conical and forced ferromagnetic phases is changed to first order by the thermal 
fluctuations. In a purely monoaxial system the pitch of the conical state remains
independent of temperature and magnetic field, as in mean field theory, even when
fluctuations are taken into account. 
However, in presence of weak Dzyaloshinskii-Moriya interactions in the plane 
perpendicular to the chiral axis, thermal fluctuations induce a dependence of
the pitch on temperature and magnetic field. This may serve to determine the
nature of magnetic interactions in such systems.
\end{abstract}

\pacs{111222-k}

\maketitle


\section{Introduction}

Much theoretical and experimental effort is being devoted to the study of monoaxial
helimagnets \cite{Han17,Yonemura17,Goncalves17,Laliena17a,Laliena16b,Shinozaki16,Nishikawa16,
Laliena16a,Masaki17,Tsuruta16,Sirica16,Garst17,
Mankovsky16,Togawa16,Kishine16,Togawa15,
Bornstein15,Dyadkin15,Kishine14,Chapman14,Ghimire13,Togawa13,Togawa12} 
due to their singular magnetic properties, which are very interesting both from the 
fundamental physics and the practical point of views. Particularly interesting are their
potential applications to spintronics.

Theoretically, the magnetic phase diagram has been extensively studied within mean field 
theory \cite{Laliena17a,Laliena16b,Shinozaki16,Nishikawa16,Laliena16b}.
However, the investigation of the effect of correlations in the thermal fluctuations, 
which may change some features of the phase diagram, has started only very recently 
\cite{Masaki17}.
In cubic helimagnets, it is known that thermal fluctuations modify the free energy of the
different states in such a way that a metastable skyrmion lattice becomes the thermodynamical
equilibrium state \cite{Muehlbauer09,Laliena17b}. 
In monoaxial helimagnets it is not expected that fluctuations cause such dramatic effects,
but the nature of the phase boundaries and some features of the equilibrium states 
can be modified. Indeed, in Ref.~\onlinecite{Masaki17} Masaki and Stamps reported an analysis,
using the Green function method, of the role of fluctuations and anisotropies in the monoaxial 
helimagnet with a magnetic field applied along the chiral axis.
They concluded that the phase boundaries and the nature of the transitions are modified.
In particular, they found metastability in the vicinity of the phase boundary and pointed out
to the possibility of a first order phase transition.

In this paper we analyze the effect of thermal fluctuations in the monoaxial helimagnet in the
presence of a magnetic field parallel to the chiral axis via a saddle point expansion.
It is shown that thermal fluctuations change the nature of the conical to forced ferromagnetic 
(FFM) phase transition from second to first order. It is also shown that the pitch of
the conical phase, which is independent of magnetic field and temperature within the standard mean
field theory, where the correlations of fluctuations
are neglected \cite{Laliena17a}, acquires a dependence on magnetic field and temperature due to
the fluctuations if chiral interactions of Dzyaloshinskii-Moriya (DM) type are present in the plane 
perpendicular to the chiral axis. However, in a purely monoaxial helimagnet, with DM interaction 
restricted to a single axis, the pitch of the conical state remains independent of magnetic field 
and temperature even if fluctuations are taken into account. 
Therefore, the dependence of the pitch on the externally imposed conditions
can be used to reveal weak magnetic interactions in monoaxial helimagnets.

The paper is organized as follows. In section~\ref{sec:model} we introduce the model and set the
notation; Sec.~\ref{sec:sp} is devoted to a description of the saddle point method used to
study the model; Sec.~\ref{sec:FFM} briefly analyzes the FFM state; Sec.~\ref{sec:CH} is devoted
to the study of the conical state; in Sec.~\ref{sec:mono} the results of the previous sections are
applied to the purely monoaxial helimagnet, and in Sec.~\ref{sec:nearly} we study the effects of weak
DM interactions in the plane perpendicular to the chiral axis; the paper ends with a brief summary
and concluding remarks in Sec. \ref{sec:conc}.

\section{Model \label{sec:model}}

We consider a classical spin system with FM and DM interactions along three perpendicular axes,
$\{\hat{x},\hat{y},\hat{z}\}$, and single ion magnetic anisotropy along an axis $\hat{u}$.
For simplicity, the FM interaction is taken isotropic in space, with strength $J$,
but the DM interaction is different along the three different axes.
In the continuum limit the energy is given by the effective Hamiltonian 
$\mathcal{H}=\epsilon_0\mathcal{W}$ where $\epsilon_0$ sets the energy scale,
and $\mathcal{W}$, a functional of the unit vector field $\hat{n}$ that represents the direction
of the local magnetic moment, can be written as the integral of a density, $\mathcal{W}=\int d^3x W$,
with
\begin{equation}
\frac{W}{q_0} = \frac{1}{2}\sum_i\partial_i\hat{n}\cdot\partial_i\hat{n}
+q_0\hat{n}\cdot\vec{D}_\rho\times\hat{n} - q_0^2\gamma(\hat{u}\cdot\hat{n})^2
-q_0^2\vec{h}\cdot\hat{n}. \label{eq:W}
\end{equation}
In the above expression $\vec{D}_\rho=\sum_i\hat{x}_i\rho_i\partial_i$ is a differential operator, 
with $\partial_i=\partial/\partial x_i$, and $x_i$ runs over $\{x,y,z\}$ in the obvious way. 
The dimensionless coefficients $\rho_i$ are real numbers that set the relative strength of the DM 
interaction along each axis.
The first term in~(\ref{eq:W}) gives the FM exchange interaction; the second term represents 
the DM interaction, whose overall strength relative to the exchange interaction is given by $q_0$, 
which has the dimensions of inverse length; the third term corresponds to the single ion
anisotropy along the axis given by the unit vector $\hat{u}$, and the last term is the
Zeeman energy. The dimensionless parameters $\gamma$ and $h$ are proportional to the strength 
of the single ion anisotropy and the applied magnetic field, respectively.

Notice that $\vec{D}_\rho$ does not transform as a vector under rotations. In covariant notation 
the DM interaction has to be written as $\rho_{ijk}n_i\partial_jn_k$, where $\rho_{ijk}$ is a tensor
antisymmetric under the exchange of $i$ and $k$, and summation over repeated indices is understood.
Nevertheless, we find it convenient to work with the non-covariant notation. Hence, the equations 
presented in this paper hold in the reference frame in which $\rho_{ijk}=\rho_i\epsilon_{ijk}$, 
where $\epsilon_{ijk}$ is the totally antisymmetric tensor.

The cubic helimagnet is obtained if $\rho_i=1$ for all $i$, and the monoaxial helimagnet 
if $\rho_x=\rho_y=0$ and $\rho_z=1$. In the latter case the magnetic anisotropy should be
directed along the same axis as the DM interaction, and therefore $\hat{u}=\hat{z}$.

The equilibrium properties of the system at temperature $T$ are given by the partition function, 
\begin{equation}
\mathcal{Z} = \int [d^2\hat{n}] \exp[-\mathcal{W}/t],
\end{equation}
where $t=T/T_0$ is a dimensionless temperature, with $T_0=\epsilon_0/k_\mathrm{B}$.

\section{Saddle point expansion \label{sec:sp}}

The dimensionless temperature, $t$, is a large number if $T\ll T_0$ and the partition 
function can be obtained by the saddle point expansion, as follows \cite{Laliena17b}. 
Let $\hat{n}_0$ be a stationary point, that is, a solution of the Euler--Lagrange equations, 
$\delta\mathcal{W}/\delta\hat{n}=0$, which read
\begin{equation}
\nabla^2\hat{n}_0 - 2q_0(\vec{D}_\rho\times\hat{n}_0) + 
2q_0^2\gamma(\hat{u}\cdot\hat{n}_0)\hat{u}+q_0^2\vec{h} = \mu\hat{n}_0, 
\label{eq:EL}
\end{equation}
where $\mu$ is a position dependent Lagrange multiplier that implements the constraint
$\hat{n}_0^2=1$, which supplements Eq.~(\ref{eq:EL}). 
Notice that the FFM state, with constant $\hat{n}_0$, is always a solution of the Euler-Lagrange
equations.

The field $\hat{n}$ in the neighborhood of $\hat{n}_0$ can be written in terms of two real
fields $\xi_\alpha$ ($\alpha=1,2$) as
\begin{equation}
\hat{n} = \sqrt{1-\xi^2}\hat{n}_0+\sum_\alpha\xi_\alpha\hat{e}_\alpha, 
\end{equation}
where the three unit vectors $\{\hat{e}_1,\hat{e}_2,\hat{n}_0\}$ form a right-handed orthonormal 
triad. They can be parametrized in terms of the two angles $\theta$ and $\psi$ 
(determined by $\hat{n}_0$) as
\begin{eqnarray}
\hat{e}_1 &=& (\cos\theta\cos\psi,\cos\theta\sin\psi,-\sin\theta), \\
\hat{e}_2 &=& (-\sin\psi,\cos\psi,0), \\
\hat{n}_0 &=& (\sin\theta\cos\psi,\sin\theta\sin\psi,\cos\theta).
\end{eqnarray}
Let us expand $\mathcal{W}$ in powers of $\xi_\alpha$ up to quadratic order:
\begin{equation}
\mathcal{W} = \mathcal{W}(\hat{n}_0) + 
\frac{q_0}{2}\int d^3x \sum_{\alpha,\beta}\xi_\alpha K_{\alpha\beta}\xi_\beta
+ O(\xi^3), \label{eq:exp}
\end{equation}
with
\begin{eqnarray}
K_{\alpha\beta} &=& -[\nabla^2 + 2W(\hat{n}_0)/q_0 + q_0^2\vec{h}\cdot\hat{n}_0]\delta_{\alpha\beta}
+\partial_i\hat{e}_\alpha\cdot\partial_i\hat{e}_\beta 
\nonumber \\
&+&q_0(\hat{e}_\alpha\cdot\vec{D}_\rho\times\hat{e}_\beta
+\hat{e}_\beta\cdot\vec{D}_\rho\times\hat{e}_\alpha)
\nonumber \\
&-& 2q_0^2\gamma(\hat{u}\cdot\hat{e}_\alpha)(\hat{e}_\beta\cdot\hat{u})
- (2\vec{G}\cdot\nabla+\nabla\cdot\vec{G})\epsilon_{\alpha\beta},
\label{eq:K}
\end{eqnarray}
where $\epsilon_{\alpha\beta}$ is the two dimensional antisymmetric unit tensor,
\begin{equation}
\vec{G}=\sum_i\left(\hat{e}_1\cdot\partial_i\hat{e}_2+q_0\rho_i\hat{x}_i\cdot\hat{n}_0
\right)\hat{x}_i,
\end{equation}
and $W(\hat{n})$ is given by Eq.~(\ref{eq:W}). 
The linear term in Eq.~(\ref{eq:exp}) vanishes on account of the Euler-Lagrange equations.

The fluctuation operator $K_{\alpha\beta}$ is a symmetric differential operator that is positive
definite if $\hat{n}_0$ is a local minimum of $\mathcal{W}$. In this case the free energy
density,
$f=-(t/V)\ln\mathcal{Z}$, can be
obtained from the saddle point method \cite{ZinnJustin97}, which is an asymptotic expansion 
in powers of $t$ that to lowest order, ignoring some irrelevant constants, gives
\begin{equation}
f = W(\hat{n}_0) + (t/V)\ln\sqrt{\det{KK_0^{-1}}} + O(t^2). \label{eq:F}
\end{equation}
The constant operator $K_{0\alpha\beta}=-\nabla^2\delta_{\alpha\beta}$ is introduced merely as
a convenient way of normalizing the contribution of fluctuations to the free energy. 
In the Quantum Field Theory jargon, the first term of~(\ref{eq:F}) is called 
the \textit{tree level} and the term proportional to $t^n$ the $n$-\textit{loop} order.
If $K$ is not positive definite the stationary point is unstable and the saddle point 
expansion does not exist.

The 1-loop term diverges in the continuum limit due to the short-distance fluctuations
and a short-distance cut-off is necessary. In solid state physics it is naturally provided 
by the crystal lattice. 
The fluctuation free energy is dominated by the short-distance fluctuations and 
depends strongly on the cut-off \cite{Muehlbauer09}.
Hence, the comparison of free energies of states computed with different cut-off schemes 
(different lattice discretization) is not meaningful. The low lying spectrum of $K$, however,
is well defined in the continuum limit and shows a weak dependence on the cut-off.

The 1-loop approximation is valid if the terms of order $\xi^3$ and higher that are 
neglected in (\ref{eq:F}) do not give a large contribution. Since the leading contribution 
of the cubic term vanishes by symmetry, the contribution of the higher order
terms relative to the quadratic terms can be estimated by the ratio 
$\langle \xi^4\rangle/\langle\xi^2\rangle\sim\langle\xi^2\rangle=t\mathrm{Tr} K^{-1}/q_0V$.

In the remaining of the paper we consider the magnetic field and the magnetic anisotropy
along the $\hat{z}$ axis, so that $\hat{h}=h\hat{z}$ and $\hat{u}=\hat{z}$. With no loss
we take $h\geq 0$. 

\section{Forced ferromagnetic state \label{sec:FFM}}

The FFM state is always a stationary point, with $\theta=0$ and $\psi$ undetermined
(may be taken as $\psi=0$). 
Its $K$ operator, 
\begin{equation}
K_{\alpha\beta} = [-\nabla^2+q_0^2(h+2\gamma)]\delta_{\alpha\beta} - 2\rho_zq_0\partial_z\epsilon_{\alpha\beta},
\end{equation}
is readily diagonalized by Fourier transform, and its spectrum reads
\begin{equation}
\lambda_\pm = k_x^2+k_y^2+(k_z\pm \rho_zq_0)^2 + q_0^2(h+2\gamma-\rho_z^2).
\end{equation}
where $\vec{k}$ is the wave vector of the eigenfunction.
The lowest eigenvalue is attained for $k_x=k_y=0$ and $k_z=\pm \rho_zq_0$ and reads 
$\lambda_{\mathrm{min}}=(h+2\gamma-\rho_z^2)q_0^2$. Therefore, the FFM state is
stable for $h>h_\mathrm{c}$ and unstable for $h<h_\mathrm{c}$, where
\begin{equation}
h_\mathrm{c} = \rho_z^2-2\gamma \label{eq:hc}
\end{equation}
is the tree level (mean field) critical field.

\section{Conical state \label{sec:CH}}

The conical state, which has the form $\theta=\theta_0$ and $\psi=qz$, where $\theta_0$ and $q$
are constants, is a stationary state for any value of the $\rho_i$.
The Euler--Lagrange equations are satisfied if and only if it holds the relation 
\begin{equation}
\cos\theta_0=\frac{h}{h_\mathrm{c}-\Delta^2(q)},
\label{eq:ct0}
\end{equation}
where
\begin{equation}
\Delta(q)=q/q_0-\rho_z. \label{eq:Delta}
\end{equation}
Since $|\cos\theta_0|\leq 1$, this stationary point exists only for
\begin{equation}
\Delta^2\leq h_c-h. \label{eq:bound}
\end{equation}
This equation sets bounds to the pitch of the conical state, $q$, and implies also
$0\leq h \leq h_\mathrm{c}$.
A second possibility for Eq.~(\ref{eq:ct0}) is 
$\Delta^2>h_\mathrm{c}+h$, which implies that the mean magnetic moment is opposite to the applied 
magnetic field, lies in the unstable region and need not be considered.

The tree level free energy of the conical state is a function of the wave vector $q$:
\begin{equation}
W_{\mathrm{C}}(\Delta) = \frac{q_0^2}{2}\left[\Delta^2 - \rho_z^2 - \frac{h^2}{h_\mathrm{c}-\Delta^2}\right].
\end{equation}
The equilibrium value of $q$ is determined by minimizing the free energy in the region where 
the stationary point is locally stable. The minimum is attained at $\Delta=0$, 
and thus the equilibrium value is $q_{\mathrm{eq}}=\rho_zq_0$, which is independent of $\rho_x$, $\rho_y$,
\textit{i.e.} of the DM interaction in the transverse plane XY, and of the magnetic field $h$
and the strength of the uniaxial anisotropy, $\gamma$.

The fluctuation operator can be readily obtained 
\begin{eqnarray}
K_{11} &=& -\nabla^2+q_0^2A, \label{eq:K11} \\
K_{22} &=& -\nabla^2, \label{eq:K22} \\
K_{12} &=& -2q_0\sin\theta_0(\rho_x\cos qz\partial_x+\rho_y\sin qz\partial_y) \nonumber \\
      &+& 2q_0\Delta\cos\theta_0\rho_z\partial_z,
\label{eq:K12}
\end{eqnarray}
where
\begin{equation} 
A = h_\mathrm{c}-\Delta^2 - h^2/(h_\mathrm{c}-\Delta^2) \\
\end{equation}
is a constant.
Notice that $A$ is positive in the neighborhood of $\Delta=0$ owing to the 
inequality~(\ref{eq:bound}). However, it is negative if $\Delta^2>h_\mathrm{c}+h$.
In appendix~\ref{app:K} it is shown that the operator $K$ is
positive definite for $\Delta=0$ if $A_0\geq 0$ and $\rho_\mathrm{m}^2<h_\mathrm{c}$, where
$A_0$ is the value of $A$ at $\Delta=0$ and $\rho_\mathrm{m}=\max\{|\rho_x|,|\rho_y|\}$. Thus the
conical state is a locally stable stationary state if the DM interaction in the plane
perpendicular to the propagation direction is weak enough.

In the remaining of the paper we restrict our attention to nearly monoaxial helimagnets, in which 
the DM interactions in the plane perpendicular to the chiral axis, $\hat{z}$, are much weaker 
than along this axis. For simplicity, we consider isotropic interactions within the perpendicular
plane, so that $\rho_x=\rho_y=\rho_\mathrm{T}\ll\rho_z$. With no loss we set $\rho_z=1$.

For a nearly monoaxial helimagnet, the 1-loop free energy of the conical state can be obtained 
perturbatively by an expansion in powers of $\rho_\mathrm{T}$.
The fluctuation operator can be written as $K=K^{(0)}+\rho_\mathrm{T}Q$, where $K^{(0)}$ corresponds to
the monoaxial helimagnet, given by setting $\rho_x=\rho_y=0$ in Eqs.~(\ref{eq:K11})-(\ref{eq:K12}),
and
\begin{equation}
Q_{\alpha\beta} = -2q_0\sin\theta_0(\cos qz\partial_x+\sin qz\partial_y) \epsilon_{\alpha\beta}.
\label{eq:Q}
\end{equation}
The 1-loop free energy can be expanded in powers of $Q$ as follows
\begin{equation}
\ln\det K = \Tr\ln K^{(0)} 
- \sum_{n=1}^\infty\frac{(-1)^n}{n}\rho_\mathrm{T}^n\Tr\left(Q{K^{(0)}}^{-1}\right)^n.
\label{eq:Qexp}
\end{equation}
Then, the free energy to 1-loop order can be written as
\begin{equation}
f(\Delta)= W_{\mathrm{C}}(\Delta) + t \left[I_0(\Delta) 
- \sum_{n=1}^\infty (-1)^n \rho_\mathrm{T}^n I_n(\Delta) \right],
\label{eq:fe}
\end{equation}
where 
\begin{equation}
I_0(\Delta) = \frac{1}{2V}\Tr\ln (K^{(0)}K_0^{-1}) \label{eq:I0}
\end{equation}
and, for $n\geq 1$,
\begin{equation}
I_n(\Delta) = \frac{1}{2nV}\Tr\left( Q{K^{(0)}}^{-1} \right)^n. \label{eq:In}
\end{equation}
These functions are studied in appendix~\ref{app:I} for $n\leq 2$.
It happens that $I_1(\Delta)$ vanishes. Some of these functions, for instance $I_0$, are
ultraviolet divergent and thus a short distance cut-off has to be introduced.
For the numerical evaluation we use a sharp cut-off in the wave vectors, 
$|\vec{k}|<\Lambda$, with $\Lambda/q_0=20$, a value appropriate for CrNb$_3$S$_6$. 

\section{Monoaxial helimagnet \label{sec:mono}}

In the previous section it has been shown that the tree level equilibrium period of the conical state
is $q_{\mathrm{eq}}=\rho_zq_0$, independent of magnetic field and the other parameters of the model. 
Indeed, the tree level free energy is an even function of $\Delta$ and thus $\Delta=0$ has to be
either a maximum or a minimum. It turns out that it is always a minimum in the region of stability
of the conical state. 
It was shown in Ref.~\onlinecite{Laliena17b} that in cubic helimagnets
the 1-loop fluctuations induce a dependence of the conical state wave vector on 
magnetic field and temperature, due to the fact that the spectrum of its fluctuation operator is not 
invariant under the change of $\Delta$ by $-\Delta$ and, therefore, the 1-loop free energy shifts the 
minimum away from $\Delta=0$. The same is expected for generic non cubic heligmagnets.

For the monoaxial helimagnet, however, the spectrum of the conical state fluctuation operator is 
invariant under the exchange of $\Delta$ by $-\Delta$, since 
$K^{(0)}_{\alpha\beta}(-\Delta)=K^{(0)}_{\beta\alpha}(\Delta)$. Thus, at least for low enough $t$,
the free energy minimum is not shifted from $\Delta=0$ and the equilibrium wave vector of the 
conical state is constant, independent of magnetic field and temperature.

Let us analyze the stability of the monoaxial helimagnet in detail.
The spectrum of $K^{(0)}$ is studied in appendix~\ref{app:K0}.
Its eigenfunctions are plane waves with wavevector $\vec{k}$ and eigenvalues
$\lambda_\sigma(\vec{k})$, with $\sigma = \pm 1$, whose expression is given in Eqs.~(\ref{eq:eval1})
and~(\ref{eq:eval2}). Hence, the spectrum of $K$ contains two branches.
The $\sigma=+1$ branch has a gap equal to $A$. The $\sigma=-1$ branch is gapless and 
corresponds to a Goldstone boson associated to the spontaneous breaking of rotational symmetry in spin 
space around the magnetic field direction. 
The presence of the Goldstone modes does not invalidate the saddle point expansion, since 
the interactions of the Goldstone modes vanish at zero momentum \cite{Burgess00},
so that the contribution of the zero mode to the higher order terms of the saddle 
point expansion vanish. Therefore the validity of the 1-loop approximation is limited by the gap $A$.
As a criterion, we consider the 1-loop approximation reliable for $t/A\lesssim 0.2$.

The Goldstone branch becomes unstable for large $\Delta$. 
To see this, notice that $\lambda_-$ is an even function of $k_z$ that for $k_x=k_y=0$ has the following 
expansion in powers of $k_z$:
\begin{equation}
\lambda_-(k_z) = \left(1-\frac{4\Delta^2\cos^2\theta_0}{A}\right) k_z^2 + O(k_z^4).
\end{equation} 
Hence, $\lambda_-$ becomes negative if $4\Delta^2\cos^2\theta_0/A>1$.
Thus, the conical state becomes unstable for $|\Delta|>\Delta_\mathrm{i}$
where $\Delta_\mathrm{i}$ is the solution of
\begin{equation}
\frac{4\Delta_\mathrm{i}^2\cos^2\theta_0}{A} = 1,
\end{equation}
where $\cos\theta_0$ and $A$ are functions of $\Delta_\mathrm{i}$.
The above equation is cubic in $\Delta_i$ and can be solved analytically, but we do not write the explicit 
solution here. It happens that $\Delta_\mathrm{i}^2<h_\mathrm{c}-h$, and thus the conical state exists as a 
stationary point for $\Delta^2<h_\mathrm{c}-h$, but it is stable only if $\Delta^2<\Delta_\mathrm{i}^2$, while
for $\Delta_\mathrm{i}^2<\Delta^2<h_\mathrm{c}-h$ is unstable.
For $h_\mathrm{c}-h<\Delta^2<h_\mathrm{c}+h$ the conical state does not exists, owing to Eq.~\ref{eq:ct0},
and for $\Delta^2>h_\mathrm{c}+h$ the conical state exists but it is unstable.
The same behavior was found in the cubic case \cite{Laliena17b}.
Obviously, the free energy of gaussian (1-loop) fluctuations is meaningful only for
$|\Delta|<\Delta_\mathrm{i}$.

\begin{figure}[t!]
\centering
\includegraphics[width=0.8\linewidth,angle=0]{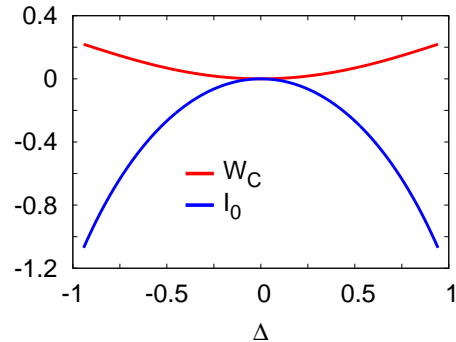}%
\caption{
Components of the free energy (tree level, $W_{\mathrm{C}}$, and 1-loop, $I_0$) as
a function of $\Delta$ for $\gamma=-2.58$ and $h=4$.
\label{fig:fecomp}
}%
\end{figure}

The free energy to 1-loop order is given by setting $\rho_\mathrm{T}=0$ in Eq.~(\ref{eq:fe}). 
It happens that $I_0(\Delta)$
is an even function of $\Delta$ that has the opposite sign of $W_\mathrm{C}(\Delta)$. 
Thus, there is a competition between
the tree level and 1-loop components of the free energy.
Fig.~\ref{fig:fecomp} displays $W_{\mathrm{C}}$ and $I_0$ as a function
of $\Delta$ for $\gamma=-2.58$ and $h=4$, with a cut-off in wave numbers $\Lambda/q_0=20$, appropriate for
CrNb$_3$S$_6$. Notice that $I_0(\Delta)$ remains finite and well defined in the limits 
$\Delta\rightarrow\pm\Delta_i$.

\begin{figure}[t!]
\centering
\includegraphics[width=0.3\linewidth,height=0.3\linewidth,angle=0]{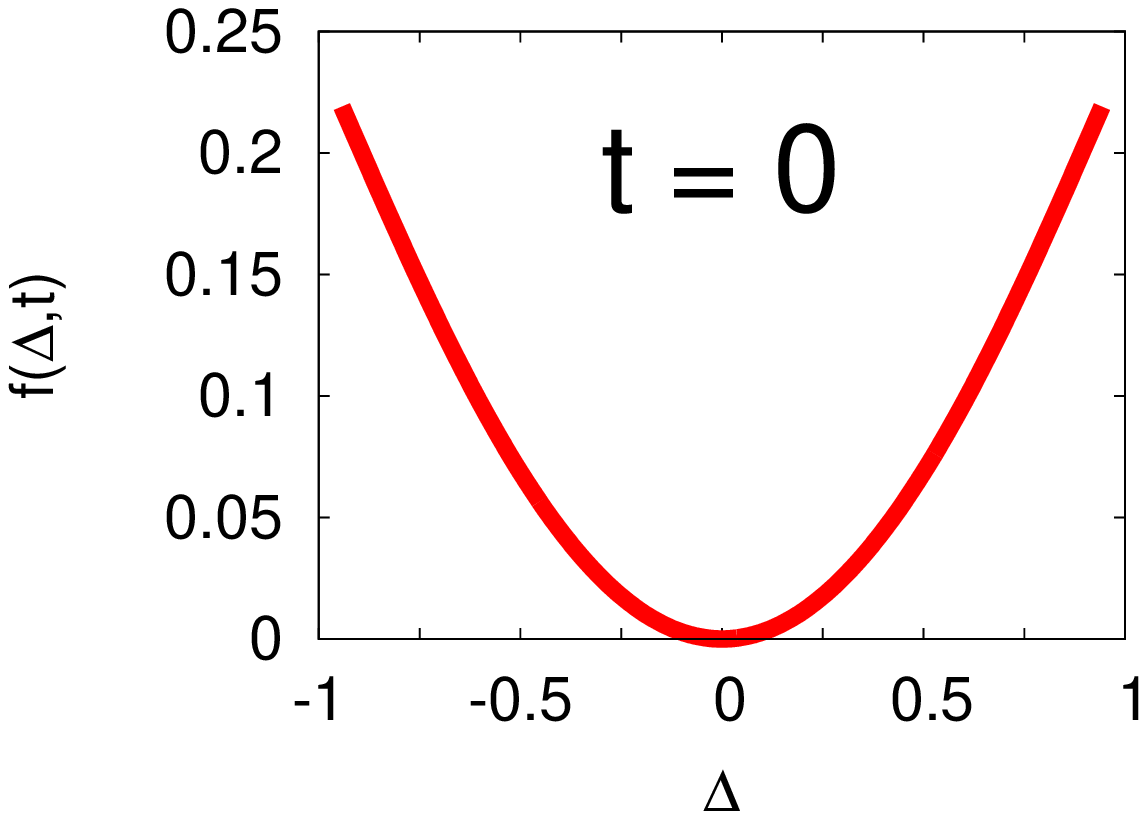}%
\includegraphics[width=0.3\linewidth,height=0.3\linewidth,angle=0]{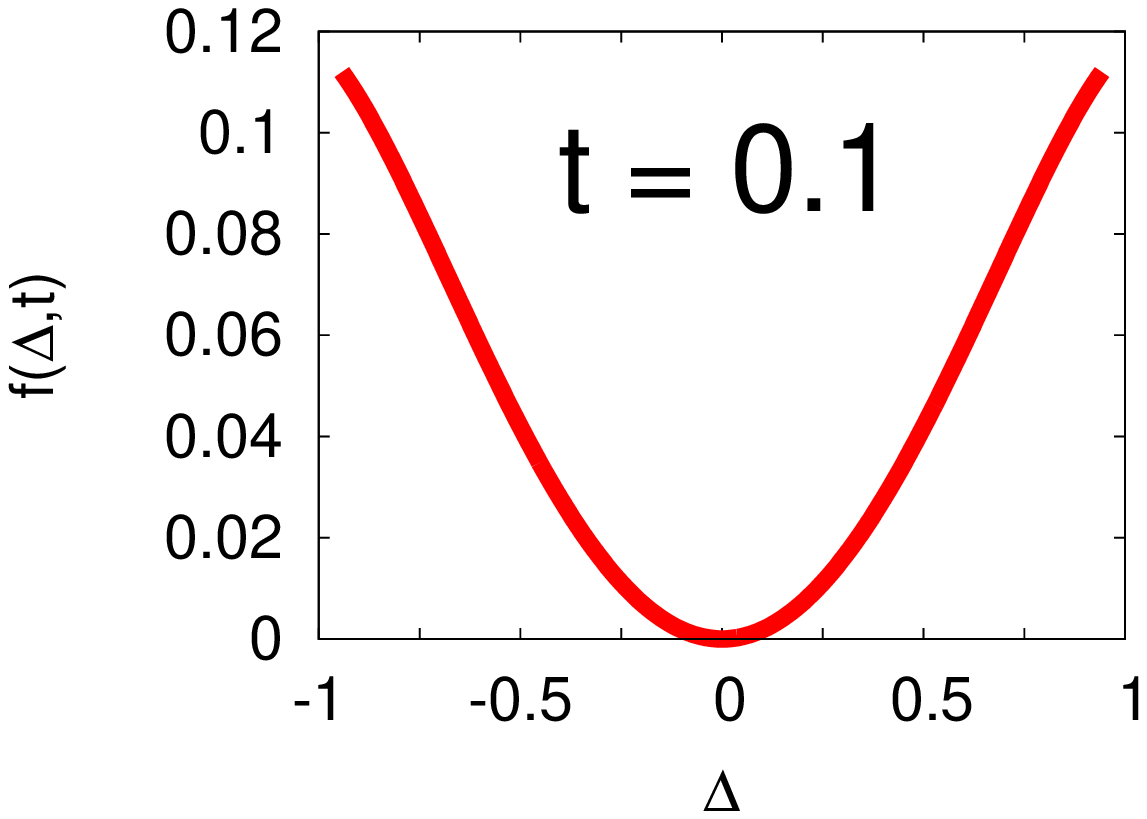}%
\includegraphics[width=0.3\linewidth,height=0.3\linewidth,angle=0]{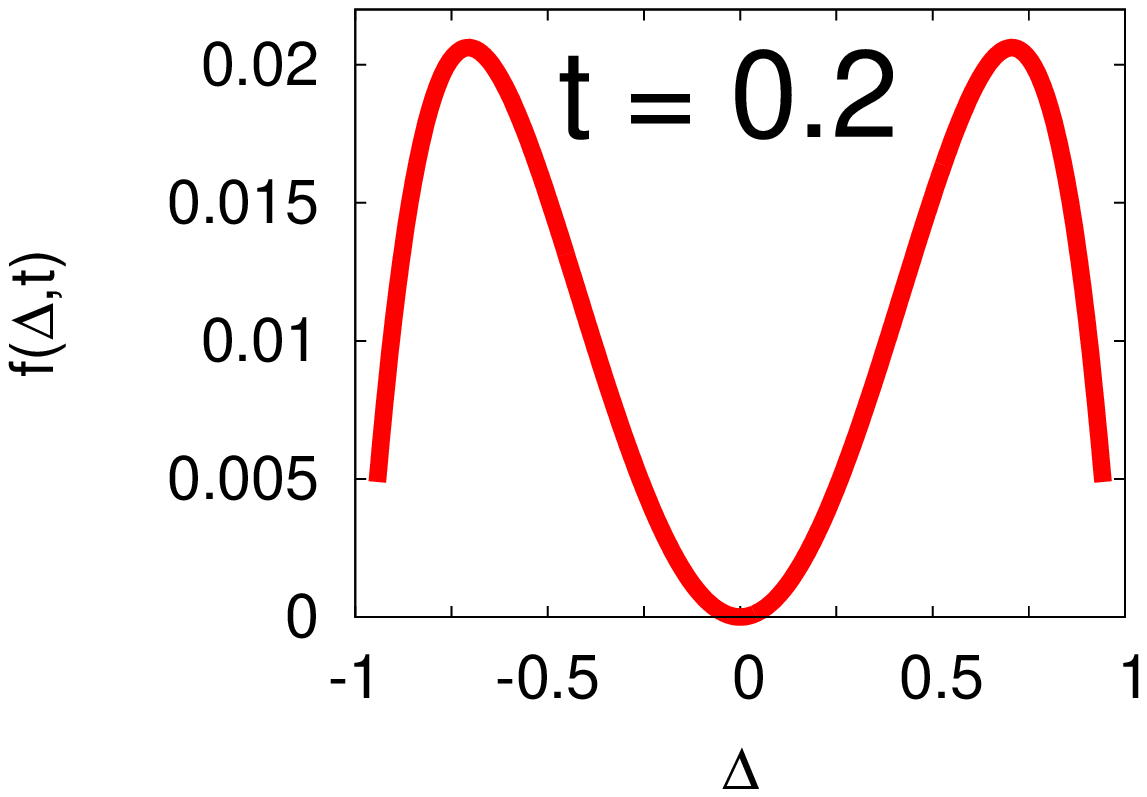}%

\includegraphics[width=0.3\linewidth,height=0.3\linewidth,angle=0]{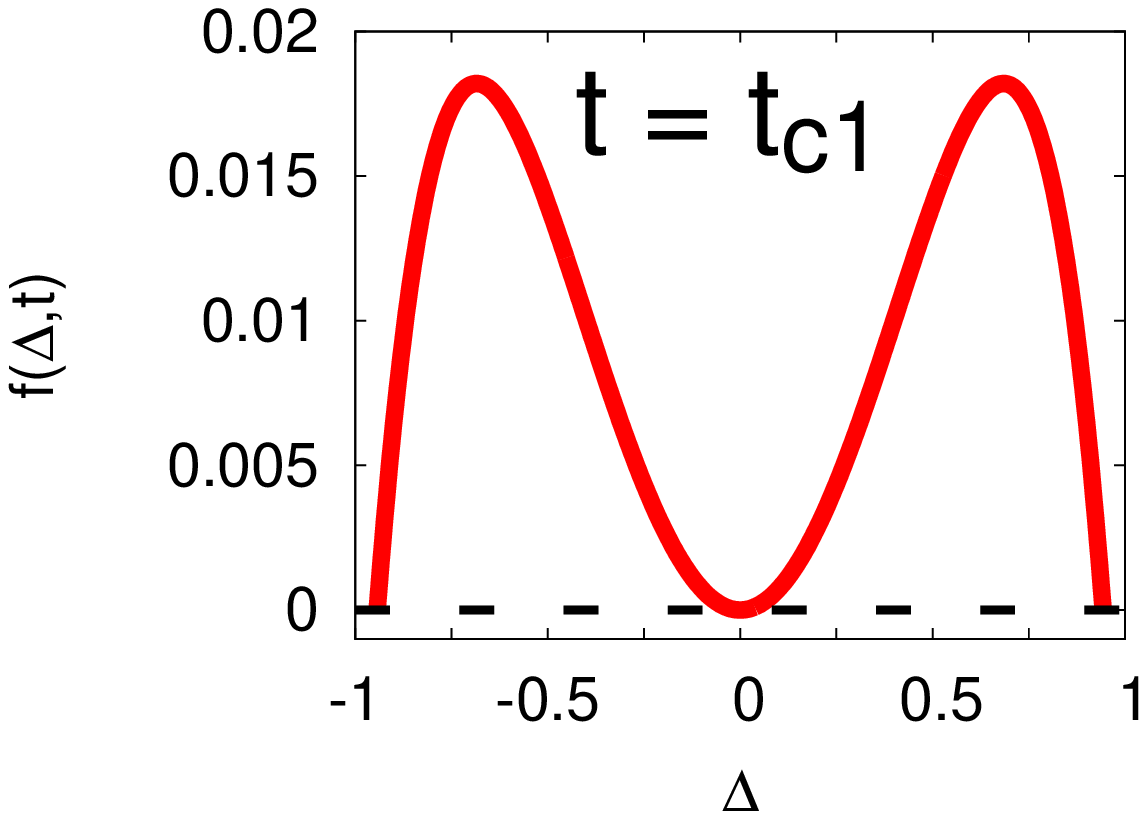}%
\includegraphics[width=0.3\linewidth,height=0.3\linewidth,angle=0]{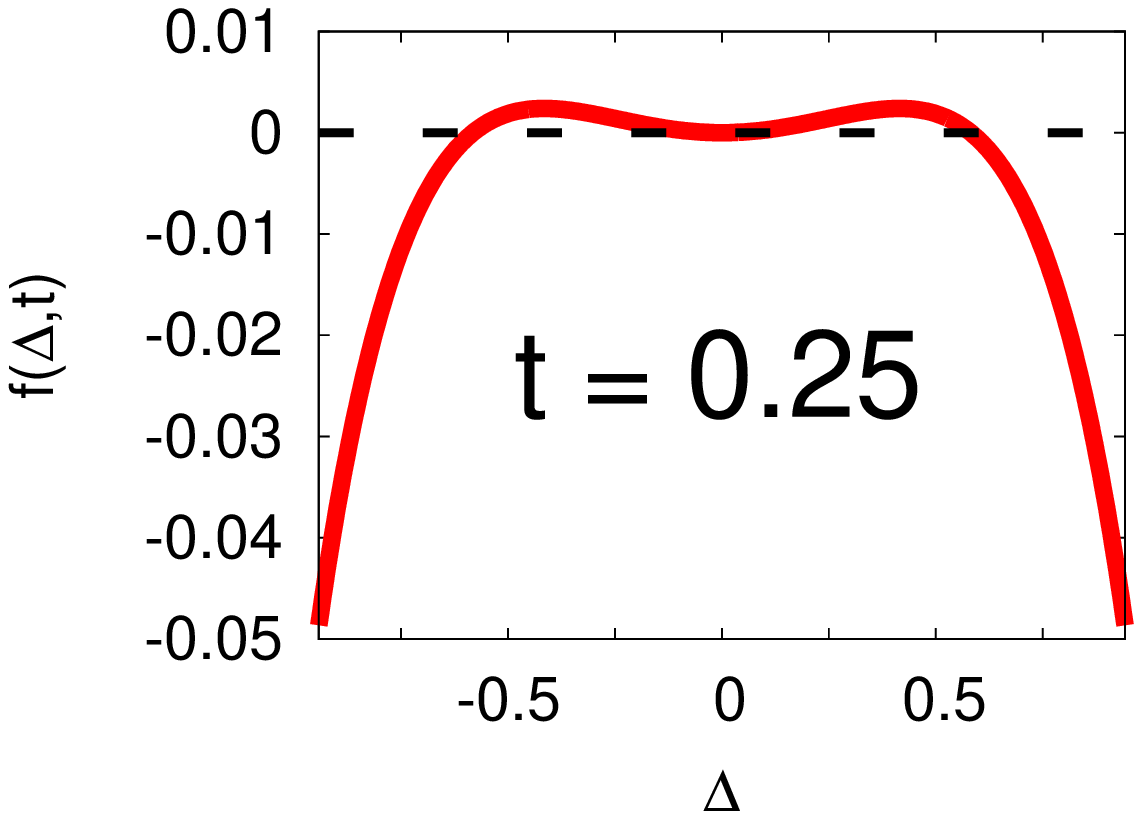}%
\includegraphics[width=0.3\linewidth,height=0.3\linewidth,angle=0]{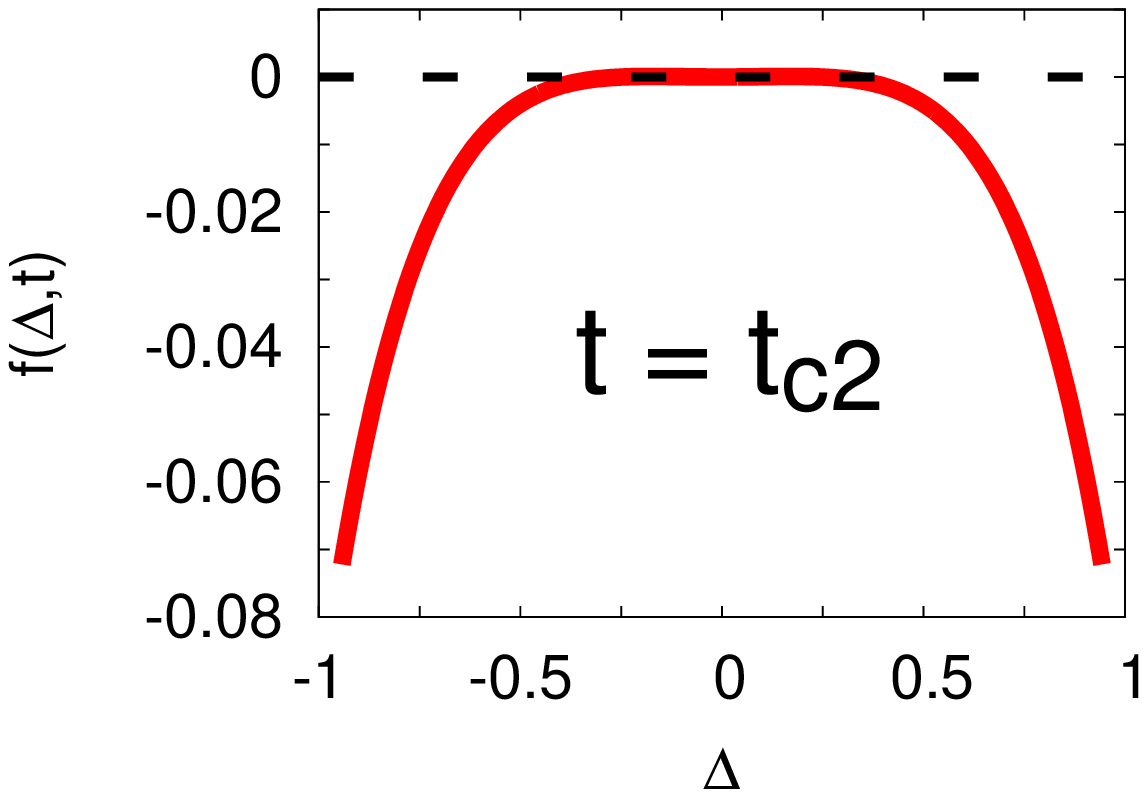}%
\caption{
The behavior of the free energy as a function of $\Delta$ for different
temperatures and fixed $h=4.0$, in the pure monoaxial case ($\rho=0$).
\label{fig:fe}}%
\end{figure}

A phase transition results from the competition between $W_\mathrm{C}$ and $I_0$.
For low $t$ the tree level dominates the free energy and its minimum is at $\Delta=0$.
At a critical temperature, $t=t_{\mathrm{c1}}$, the local minimum at $\Delta=0$ equals the free 
energy at the limiting value $\Delta_i$:
\begin{equation}
t_{\mathrm{c1}} = -\frac{W_{\mathrm{C}}(0)}{I_0(\Delta_i)}.
\end{equation}
A first order phase transition takes place at $t_{\mathrm{c1}}$.
The conical state remains metastable for $t_{\mathrm{c1}}<t<t_{\mathrm{c2}}$, where
\begin{equation}
t_{\mathrm{c2}} = -\frac{W^{\prime\prime}_{\mathrm{C}}(0)}{I^{\prime\prime}_0(0)}
\end{equation}
is the temperature at which $\Delta=0$ becomes a maximum of the free energy.
For $t>t_{\mathrm{c2}}$ the conical state is not even metastable. 
In our numerical example we obtained $t_{\mathrm{c1}}=0.2045$ and $t_{\mathrm{c2}}=0.272$.
The behavior of the free energy by increasing temperature is illustrated in Fig.~\ref{fig:fe} 
for the case $h=4.0$. 

\begin{figure}[t!]
\centering
\includegraphics[width=\linewidth,angle=0]{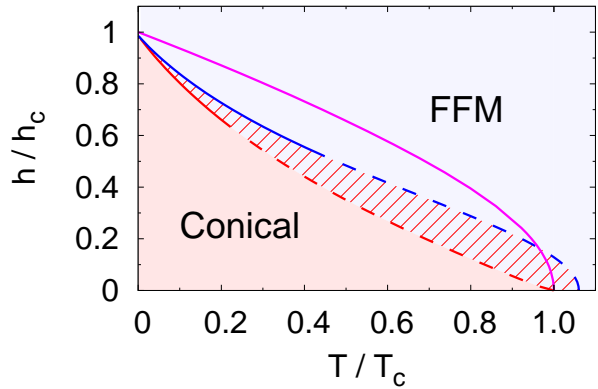}%
\caption{Phase diagram for the monoaxial helimagnet in the gaussian approximation.
The red line and blue lines corresponds to $t_{\mathrm{c1}}$ (first order phase transition)
and $t_{\mathrm{c2}}$ (appearance of the metastable conical state), respectively. 
On the broken lines the 1-loop (gaussian) approximation is not reliable.
The conical state is metastable in the region filled with red stripes.
The pink line is the phase boundary predicted by the variational mean field approximation. 
The transition in this case is of second order instability type.
The temperature is normalized to $T_\mathrm{C}$, which is the zero field
transition temperature.
\label{fig:phd}
}%
\end{figure}

The phase diagram is displayed in Fig.~\ref{fig:phd}. 
The temperature is normalized by the zero field critical temperature, $T_\mathrm{C}$.
With our choice of parameters for the numerical computations we have $T_\mathrm{C} = 0.978\,T_0$.
The red line represents the transition line, given by $t_{\mathrm{c1}}$. The
conical state is metastable in the region filled by red stripes, and disappears on the
blue line, $t_{\mathrm{c2}}$. With the criterion that the saddle point expansion is valid if 
$t/A\lesssim 0.2$, the solid lines of the $t_{\mathrm{c1}}$ and $t_{\mathrm{c2}}$ boundaries are reliable.
The broken lines may receive important contributions from higher order terms and are not reliable.

The pink line signals, for comparison, the phase boundary obtained with the 
variational mean field approach, which predicts a second order instability type 
phase transition \cite{Laliena17a}. The saddle point expansion is reliable at low temperature, 
but fails at high temperature.
The variational mean field theory is the lowest order term of a cumulant expansion, and 
neglects the correlations between the spin fluctuations at different sites. No small parameter
justify this expansion and is thus questionable, although it is more reliable at higher temperatures,
since the correlations between fluctuations diminish as temperature increses. The exception, of course,
is the zero field critical point, where the fluctuations are strongly correlated.
The conclusion is then that the phase transition is of first order at low temperature, as predicted
by the saddle point expansion, and of second order instability type at high temperature, as predicted
by the variational mean field theory. These two transitions of different nature have to be separated 
by a tricritical point. Thus the phase diagram obtained in Ref.~\onlinecite{Laliena17a} from the
variational mean field theory has to be modified at low temperature.

\section{Nearly monoaxial helimagnet \label{sec:nearly}}

In this section it is shown that the equilibrium period of the conical state of a nearly monoaxial 
helimagnet shows a weak dependence on magnetic field and temperature, proportional to 
$\rho_\mathrm{T}^2$, since thermal fluctuations
at the 1-loop level shift the free energy minimum away from $\Delta=0$.
Thus, any variation of the conical state period with temperature or field of a presumed monoaxial 
helimagnet reveals weak chiral interactions in the plane perpendicular to the chiral axis.

The equilibrium value of  $\Delta$, denoted by $\Delta_{\mathrm{eq}}$, corresponds to the minimum of the 
free energy, so that it obeys the equation
\begin{equation}\label{minimunEq}
W_{\mathrm{C}}^\prime(\Delta_{\mathrm{eq}})+t\left[I^\prime_0(\Delta_{\mathrm{eq}})
-\rho_\mathrm{T}^2I^\prime_2(\Delta_{\mathrm{eq}})\right]=0,
\end{equation}
where the prime stands for the derivative with respect to $\Delta$. 

\begin{figure}[t!]
\centering
\includegraphics[width=0.8\linewidth,angle=0]{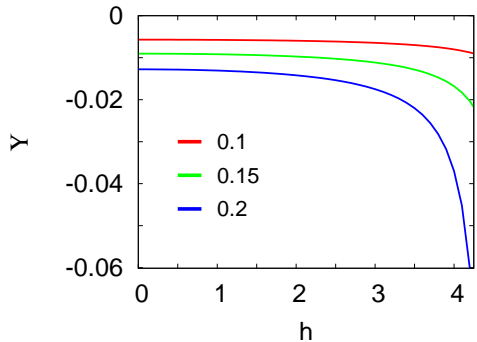}
\caption{
The wave number variation with $h$, given by the function $\Upsilon(t,h)$, for the values
of the dimensionless temperature, $t$, displayed in the legend. 
\label{fig:upsilon}}
\end{figure}

Notice that $W_{\mathrm{C}}$ and $I_0$ are even functions of $\Delta$, and it has been shown in the 
previous section that the free energy minimum is always at $\Delta=0$ if $\rho_\mathrm{T}=0$.
For small $\rho_\mathrm{T}$ the equilibrium value of $\Delta$ will be of order $\rho_\mathrm{T}^2$
and can be expressed as 
\begin{equation}
\Delta_{\mathrm{eq}} = \rho_\mathrm{T}^2 \Upsilon(t,h). \label{eq:deltaeq}
\end{equation}
Expanding Eq.~(\ref{minimunEq}) around $\Delta_{\mathrm{eq}}=0$ we get,
\begin{equation}
\Upsilon(t,h)=\frac{t I^\prime_2(0)}{W_\mathrm{C}^{\prime\prime}(0) + t I_0^{\prime\prime}(0)}.
\label{eq:upsilon}
\end{equation}
The equilibrium wave number of the conical state is given by
\begin{equation}
\frac{q_{\mathrm{eq}}}{q_0} = \rho_z + \rho_\mathrm{T}^2 \Upsilon(t,h). 
\end{equation}
The function $\Upsilon(t,h)$ is plotted as a function of $h$ for several values of $t$ in
Fig.~\ref{fig:upsilon}.
Notice that $\Upsilon(t,h)$ is negative and decreases with $h$. Thus the wave vector decreases
(and the period increases) with temperature and magnetic field. This is consistent with the fact 
that the FFM state will be attained by incresing temperature or magnetic field.

\section{Conclusions \label{sec:conc}}

Mean field theory, which neglects the correlations between thermal fluctuations,
predicts a second order instability type phase transition between the conical and the FFM
states in a monoaxial helimagnet with a magnetic field parallel to the chiral axis.
In this paper we have shown that the correlation of fluctuations at gaussian level, computed via
the saddle point expansion to 1-loop order, changes the nature of the phase transition 
from second to first order.
Signals of a first order transition for this system have also been noticed in 
Ref.~\onlinecite{Masaki17}, where the correlation of fluctuations are included via the Green's 
function method.
The saddle point expansion, which is an asymptotic expansion in powers of $T/T_0$
and therefore not valid at high temperature, is more reliable at low temperature than the 
variational mean field theory. Therefore, it is likely that the transition changes from first to 
second order as temperature increases. This means that a tricritical point appears on the phase 
boundary.

It is worthwhile to point out that fluctuations of different types drive phase transitions to first 
order in different systems, as superconductors and liquid crystals \cite{Halperin74} and
cubic helimagnets at low magnetic field \cite{Brazovskii75,Janoschek13}.
 
According to mean field theory, the pitch of the conical state of an helimagnet is independent
of temperature and magnetic field. For a pure monoaxial helimagnet, with DM interactions only 
along one axis, the correlated fluctuations preserve this feature of the conical state.
But in the presence of DM interations in the plane perpendicular to the chiral axis, however week
they are, thermal fluctuations induce a dependence of the pitch on temperature and magnetic field.
This variation of the pitch with the externally imposed conditions may be used to determine the
nature of the chiral interactions in helimagnets.
Thus, any temperature or field dependence of the pitch in presumed monoaxial chiral magnets,
such as CrNb$_3$S$_6$, can be interpreted as a departure from a purely monoaxial DM interaction.

\begin{acknowledgments}
The authors thank Y. Masaki for frutiful discussions in the early stage of the collaboration.
Grant No. MAT2015-68200- C2-2-P from the Spanish Ministry of Economy and Competitiveness,
Grant No. 25220803 from the scientific JSPS Grant-in-Aid for Scientific Research (S),
and Grant Number JP17H02923 from JSPS KAKENHI are acknowledged.
This work was also supported by the MEXT program for promoting the enhancement of research 
universities, by the JSPS Core-to-Core Program, A. (Advanced Research Networks), by the 
Chirality Research Center (Crescent) in Hiroshima University, and by JSPS and RFBR under the 
Japan - Russia Research Cooperative Program. 
\end{acknowledgments}

\appendix

\section{Local stability of the conical state \label{app:K}}

Let us show that the conical state fluctuation operator, $K$, defined by 
Eqs.~(\ref{eq:K11})-(\ref{eq:K12}), with $\Delta=0$ is definite positive if $\rho_x$
and $\rho_y$ are small enough (here we do not assume isotropy in the plane perpendicular 
to the chiral axis). The most general square integrable wave function can be written as
\begin{equation}
\xi_\alpha(\vec{x}) = \int\frac{d^3k}{(2\pi)^3}
\exp(\mathrm{i}\vec{k}\cdot\vec{x})\tilde{\xi}_\alpha(z,\vec{k}),
\end{equation}
where $-q_0/2\leq k_z\leq q_0/2$ and $\tilde{\xi}_\alpha(z,\vec{k})$ is periodic in $z$: 
\begin{equation}
\tilde{\xi}_\alpha(z+L_0)=\tilde{\xi}_\alpha(z), 
\end{equation}
with $L_0=2\pi/q_0$.
The expectation value of $K$ with this generic wave function is given by
\begin{equation}
\bra{\xi}K\ket{\xi} = \int\frac{d^3k}{(2\pi)^3}\int_0^{L_0} dz T(z,\vec{k}),
\end{equation}
with
\begin{eqnarray}
&&T(z,\vec{k}) = \sum_\alpha\left[ k^2|\tilde{\xi}_\alpha|^2+|\tilde{\xi}_\alpha^\prime|^2
+2\mathrm{Im}(k_z\tilde{\xi}_\alpha^*\tilde{\xi}_\alpha^\prime)
\right] + q_0^2A_0|\tilde{\xi}_1|^2 \nonumber \\
&& + 2\mathrm{Im}[q_0\sin\theta_0(\rho_xk_x\cos q_0z + \rho_yk_y\sin q_0z)
\tilde{\xi}_1^*\tilde{\xi}_2],
\end{eqnarray}
where $A_0=h_c-h^2/h_c$ is the value of $A$ at $\Delta=0$, the prime stands for derivative with respect 
to $z$, and we omite the arguments $z$ and $\vec{k}$ in the functions $\tilde{\xi}_\alpha$ 
and $\tilde{\xi}_\alpha^\prime$. Using the inequalities
$a+b\geq a-|b|$, valid for any pair of real numbres $a$ and $b$, $|\mathrm{Im}\,c|\leq|c|$, valid for any
complex number $c$, and
\begin{equation}
|\rho_xk_x\cos q_0z + \rho_yk_y\sin q_0z|\leq\rho_{\mathrm{m}}k_\mathrm{T},
\end{equation}
where $\rho_{\mathrm{m}}=\max\{|\rho_x|,|\rho_y|\}$ and $k_\mathrm{T}^2=k_x^2+k_y^2$, we have
\begin{equation}
\bra{\xi}K\ket{\xi} \geq \int\frac{d^3k}{(2\pi)^3}\int_0^{L_0} dz \left[
\sum_\alpha\left( |k_z\tilde{\xi}_\alpha|-|\tilde{\xi}_\alpha^\prime| \right)^2 + \Pi\right],
\end{equation}
where
\begin{equation}
\Pi = k_\mathrm{T}^2|\tilde{\xi}_2|^2 
+ (k_\mathrm{T}^2+q_0^2A_0)|\tilde{\xi}_1|^2 
- 2q_0\sin\theta_0\rho_{\mathrm{m}}k_\mathrm{T}|\tilde{\xi}_1||\tilde{\xi}_2|
\end{equation}
is a quadratic form in $|\tilde{\xi}_\alpha|$. The operator $K$ will be positive definite
if $\Pi$ is positive definite for any $k_\mathrm{T}$. The condition for the quadratic form
to be positive definite is that all its principal minors be positive, that is
\begin{eqnarray}
k_\mathrm{T}^2 + q_0^2A_0 \geq 0, \\
k_\mathrm{T}^2(k_\mathrm{T}^2 + q_0^2A_0 - q_0^2\rho_\mathrm{m}^2\sin^2\theta_0)\geq 0.
\end{eqnarray}
The first inequality implies $A_0\geq 0$, and the second inequality gives
$\rho_\mathrm{m}^2 \leq A_0/\sin^2\theta_0$. Recalling the expressions for $A_0$ 
and $\sin^2\theta_0$ we get
\begin{equation}
\rho_\mathrm{m}^2 \leq \rho_z^2 - 2\gamma.
\end{equation}

\section{Spectrum of $K^{(0)}$ \label{app:K0}}

The operator $K^{(0)}$, given by Eqs.~(\ref{eq:K11})-(\ref{eq:K12}) with $\rho_x=\rho_y=0$,
can be diagonalized by Fourier transform. Its normalized eigenfunctions are plane waves
\begin{equation}
\ket{\xi^\sigma(\vec{k})}=\frac{1}{\sqrt{V}}\left(\begin{array}{c}\phi_1^\sigma(\vec{k})\\ 
\phi_2^\sigma(\vec{k}) \end{array}\right) e^{i\vec{k}\cdot\vec{x}}, 
\label{eq:evec}
\end{equation}
with $\sigma=\pm 1$, $V$ is the volume and
\begin{equation}
\phi_1^{\sigma *}(\vec{k})\phi_1^{\sigma'}(\vec{k})+\phi_2^{\sigma *}(\vec{k})\phi_2^{\sigma'}(\vec{k})
=\delta_{\sigma\sigma'}.
\label{eq:normphi}
\end{equation}
They form a complete set.
The corresponding eigenvalues are 
\begin{equation}
\lambda_{\sigma}(\vec{k})=k^2+f^{(\sigma)}(k_z),
\label{eq:eval1}
\end{equation}
with
\begin{equation}
f^{(\sigma)}(k_z) = \frac{q_0^2A}{2}
\left(1+\sigma\sqrt{1+\frac{16 \Delta^2 \cos^2\theta_0}{A^2}\frac{k_z^2}{q_0^2}} \,\right).
\label{eq:eval2}
\end{equation}
The $\sigma=+1$ branch of the spectrum has a gap of value $q_0^2A$.
The $\sigma=-1$ branch is gapless and corresponds to a Goldstone boson associated to the spontaneous breaking
of the rotational symmetry in spin space corresponding to the rotation around the magnetic field direction.

The polarization of the plane waves can be chosen of the form
\begin{equation}
\left(\begin{array}{c}
\phi_1^\sigma(k_z) \\ 
\phi_2^\sigma(k_z) \end{array}\right) = \frac{1}{\sqrt{1+\Theta^2}}
\left(\begin{array}{c} 
(-\mathrm{i}\Theta)^{\frac{1-\sigma}{2}} \\ 
(-\mathrm{i}\Theta)^{\frac{1+\sigma}{2}} \\ 
\end{array}\right),
\label{eq:pol}
\end{equation}
with
\begin{equation}
\Theta(k_z,\Delta)=\frac{\left(q_0^2A^2+16\Delta^2\cos^2\theta_0 k_z^2\right)^{1/2}-Aq_0}{4\Delta\cos\theta_0 k_z}.
\end{equation}
Notice that $\lim_{\Delta\rightarrow 0}\Theta(k_z,\Delta) = 0$.

The spectrum of $K^{(0)}$ depends on $\Delta$ through $\Delta^2$ and is thus invariant under the
the exchange of $\Delta$ by $-\Delta$.

\section{The functions $I_n(\Delta)$ \label{app:I}}

The evaluation of the $I_n(\Delta)$ functions defined by Eqs.~(\ref{eq:I0}) and~(\ref{eq:In})
involves integrals over the wave number $\vec{k}$ that are
ultraviolet divergent and thus a short distance cut-off, $\Lambda$, has to be introduced. The cut-off, 
of course, is naturally provided by the underlying crystal lattice. We find it convenient to use a 
sharp cylindrical cut-off,
so that the spectrum of $K^{(0)}$ is limited to the wave vector region defined by
$|\vec{k}_\mathrm{T}|<\Lambda$ and $|k_z|<\Lambda$, where $\vec{k}_\mathrm{T}=k_x\hat{x}+k_y\hat{y}$ is the
wave vector projection onto the plane perpendicular to the magnet axis.
This cut-off choice is not unreasonable since we are dealing with monaxial helimagnets.


The function $I_0(\Delta)$ is given by the following integral: 
\begin{equation}
I_0(\Delta) = \frac{1}{2}\sum_{\sigma=\pm 1}
\int\frac{d^3k}{(2\pi)^3}\ln\left(\frac{\lambda_\sigma(\vec{k})}{k^2}\right).
\label{eq:I0ass}
\end{equation}
The integral in $\vec{k}_\mathrm{T}$ can be readily performed, and it remains an integral in $k_z$ that 
can be performed numerically. The integral is linearly divergent with the cut-off, and its leading 
term as $\Lambda\rightarrow\infty$ is
\begin{equation}
I_0(\Delta) \sim \frac{q_0^3}{8\pi^2}\left[(\ln 2+\pi/2)A + \pi\Delta^2\cos^2\theta_0 
\right]\frac{\Lambda}{q_0}.
\end{equation}
The function $\Upsilon(t,h)$ involves $I_0^{\prime\prime}(0)$, which, to leading order in $\Lambda$ 
can be obtained from (\ref{eq:I0ass}):
\begin{equation}
I_0^{\prime\prime}(0) \sim -\frac{q_0^3}{4\pi^2}\left[\ln 2+\frac{\pi}{2} 
+ \left(\ln 2 + \frac{3\pi}{2}\right)\frac{h^2}{h_\mathrm{c}^2} \right]\frac{\Lambda}{q_0}.
\end{equation}


To evaluate $I_1$ and $I_2$ we need the matrix elements of $Q$ between the eigenstates of $K^{(0)}$.
They read
\begin{eqnarray}
&&\bra{\xi^{\sigma^\prime}(\vec{k}^\prime)} Q \ket{\xi^{\sigma}(\vec{k})} =  \nonumber \\
&&-\mathrm{i}C\sum_{\alpha\beta}\epsilon_{\alpha\beta}
{\phi_\alpha^{\sigma^\prime}(k_z^\prime)}^* \phi_\beta^{\sigma}(k_z)
[k_-\delta_{\vec{k}^\prime,\vec{k}+q\hat{z}} + k_+\delta_{\vec{k}^\prime,\vec{k}-q\hat{z}}], \qquad
\label{eq:mel}
\end{eqnarray}
where $C=q_0\sin\theta_0$ and $k_\pm=k_x\pm\mathrm{i}k_y$.
The diagonal elements vanish since $q>0$ in the region of stability of the conical state.
Therefore, $I_1(\Delta)=0$.

Inserting a resolution of the identity in terms of the eigenvalues of $K^{(0)}$ into the definition
of $I_2(\Delta)$, we get
\begin{equation}
I_2(\Delta) = \frac{1}{4\rho_\mathrm{T}^2V}\sum_{\sigma\sigma^\prime}\sum_{\vec{k},\vec{k}^\prime}
\frac{|\bra{\xi^{\sigma^\prime}(\vec{k}^\prime)} Q \ket{\xi^{\sigma}(\vec{k})}|^2}
{\lambda_{\sigma^\prime}(\vec{k}^\prime)\lambda_\sigma(\vec{k})}.
\end{equation}
Using (\ref{eq:mel}) and taking the infinite volume limit we get
\begin{equation}
I_2(\Delta) = -\frac{C^2}{4}
\sum_{\sigma=\pm 1}\int\frac{d^3k}{(2\pi)^3}k_\mathrm{T}^2
\Tr[G(\vec{k})\tau_yG(\vec{k}+\sigma q\hat{z})\tau_y], \label{eq:I2int}
\end{equation}
where $\tau_y$ is the Pauli matrix and the matrix $G$ is defined by
\begin{equation}
G_{\alpha\beta}(\vec{k}) = \sum_{\sigma=\pm 1}
\frac{\phi_\alpha^\sigma(k_z)\phi_\beta^{\sigma}(k_z)^*}{\lambda_\sigma(\vec{k})}.
\end{equation}
The integral (\ref{eq:I2int}) is complicated, but actually we are only interested in the
derivative of $I_2(\Delta)$ at $\Delta=0$, which enters the function $\Upsilon(t,h)$,
and this is much simpler. It is not difficult to see that all the $\Delta$ dependence
in the right-hand side of Eq.~(\ref{eq:I2int}) is through $\Delta^2$, except for the dependence
via $q=q_0(\rho_z+\Delta)$. Hence, we only need to compute the derivative of $G_{\alpha\beta}$
with respect to $k_z$. Taking into account the relations 
\begin{eqnarray}
&&\lim_{\Delta\rightarrow 0} G_{11} = (k^2+q_0^2A_0)^{-1}, \\
&&\lim_{\Delta\rightarrow 0} G_{22} = k^{-2}, \\
&&\lim_{\Delta\rightarrow 0} G_{12} = 0, \\
&&\lim_{\Delta\rightarrow 0} \partial G_{11}/\partial k_z = -2k_z(k^2+q_0^2A_0)^{-2}, \\
&&\lim_{\Delta\rightarrow 0} \partial G_{22}/\partial k_z = -2k_z k^{-4}, \\
&&\lim_{\Delta\rightarrow 0} \partial G_{12}/\partial k_z = 0,
\end{eqnarray}
where $A_0$ is the value of $A$ at $\Delta=0$, we obtain
\begin{widetext}
\begin{equation}
I^\prime_2(0) = \frac{q_0^3A_0}{4h_c}
\int\frac{d^3k}{(2\pi)^3}k_\mathrm{T}^2
\left[\frac{1}{k^2+q_0^2A_0}\frac{2(k_z-q_0)}{[(\vec{k}-q_0\hat{z})^2]^2} 
+\frac{1}{k^2}\frac{2(k_z-q_0)}{[(\vec{k}-q_0\hat{z})^2+q_0^2A_0]^2}
- (q_0\rightarrow -q_0) \right].
\label{eq:I2pp0}
\end{equation}
\end{widetext}
The integral over $\vec{k}_\mathrm{T}$ can be readily performed and 
it remains the following integral over $k_z$:
\begin{eqnarray}
I^\prime_2(0) &=& \frac{q_0^3A_0}{8\pi^2h_c}
\int_{-\Lambda}^\Lambda dk_z (k_z-q_0)
\left\{J[k_z^2+q_0^2A_0,(k_z-q_0)^2] \right. \nonumber \\
&+& \left. J[k_z^2,(k_z-q_0)^2+q_0^2A_0]\right\},
\label{eq:I2pp0b}
\end{eqnarray}
where
\begin{eqnarray}
J(E_1,E_2) &=& \frac{E_1}{(E_2-E_1)^2}\ln\frac{E_1(\Lambda^2+E_2)}{E_2(\Lambda^2+E_1)} \nonumber \\
&+& \frac{\Lambda^2}{(E_2-E_1)(\Lambda^2+E_2)}.
\end{eqnarray}
The function $J(E_1,E_2)$ is analytic at $E_1=E_2$. The integral~(\ref{eq:I2pp0b}) is ultraviolet
finite. The logarithmic divergence coming from the integration region $k_z\sim\Lambda$ which
cancels exactly with the divergence coming from the $k_z\sim -\Lambda$ region.
It has been evaluated numerically using for $\Lambda$ the same value as in the computation of
$I_0(\Delta)$ and $I^{\prime\prime}_0(0)$.

\bibliography{references}

\end{document}